\documentclass[twocolumn,amssymb,nobibnotes,showpacs,superscriptaddress,aps,pr,tightenlines]{revtex4-1}

\usepackage[pdftex]{hyperref}
\hypersetup{colorlinks,
			linkcolor={blue!75!black!80!yellow},
			citecolor={blue!75!black!80!yellow},
			urlcolor={blue!75!black!80!yellow},
			pdfstartview=FitH}
\usepackage{graphicx}
\usepackage{mathrsfs}
\usepackage{amsthm}
\usepackage{physics}
\usepackage{xspace}
\usepackage{braket}
\usepackage{xr}
\usepackage{gensymb} 
\usepackage{xcolor,soul}
\usepackage{stmaryrd} 
\usepackage[UKenglish]{babel}
\usepackage{placeins} 
\usepackage{siunitx}
\DeclareSIUnit[number-unit-product=]\percent{\char`\%} 
\usepackage{makecell}
\usepackage{diagbox}
\usepackage{float}
\usepackage{amsmath} 
\usepackage{empheq} 

\usepackage{txfonts}  
\usepackage{txfontsb} 

\newcommand*\diff{\mathop{}\mathrm{d}}

\renewcommand{\Re}{\operatorname{Re}}
\renewcommand{\Im}{\operatorname{Im}}

\newcommand{\appropto}{\mathrel{\vcenter{
			\offinterlineskip\halign{\hfil$##$\cr
				\propto\cr\noalign{\kern2pt}\sim\cr\noalign{\kern-2pt}}}}}

\newcommand{\JL}{J_{\mathrm{L}}}
\newcommand{\JR}{J_{\mathrm{R}}}

\makeatletter
\newcommand*{\addFileDependency}[1]{
  \typeout{(#1)}
  \@addtofilelist{#1}
  \IfFileExists{#1}{}{\typeout{No file #1.}}
}
\makeatother

\usepackage{scalerel}

\usepackage{textcomp} 
\usepackage{xifthen}
\usepackage{etoolbox}
\newboolean{togglecomments}
\newboolean{togglechanges} 

\setboolean{togglecomments}{true}
\setboolean{togglechanges}{false}

\newcommand{\comment}[2]{%
    \ifbool{togglecomments}%
    {\textcolor{blue!70!black}{\small\textsf{%
    \textsuperscript{\textsc{\textsf{\MakeLowercase{#1}}}}%
    [#2]}}} 
    {}}     
\newcommand{\swap}[2]{\ifbool{togglechanges}
    {#2}  
    {\textcolor{red!70!black}{[#1]}\textrightarrow{}\textcolor{green!50!black}{[#2]}}}
\newcommand{\remove}[1]{\ifbool{togglechanges}
    {}    
    {\textcolor{red!70!black}{#1}}}
\newcommand{\inset}[1]{\ifbool{togglechanges}
    {#1}  
    {\textcolor{green!50!black}{#1}}}
\newcommand{\optional}[1]{\ifbool{togglechanges}
    {#1}  
    {\textcolor{yellow!50!orange!80!gray}{#1}}}
\newcommand{\citeremind}[1]{%
    [\textcolor{blue!75!black!80!yellow}{
        $\blacksquare$%
           \ifthenelse{\isempty{#1}}
               {}
               {\textsuperscript{\textsf{#1}}}%
        }]\xspace}
\newcommand{\todo}[1]{
    \textcolor{orange!80!yellow!95!black}{\textbf{[}%
        \ifthenelse{\isempty{#1}}%
        {\text{$\blacksquare$}}%
        {{\small\textsf{#1}}}%
        \textbf{]}}}
        
\usepackage{color,graphicx,amsfonts,amssymb,amsmath,hyperref,footmisc,epsfig,fancyhdr}

\begin{document}

\title{Non-Abelian Gauge Effect for 2D Non-Hermitian Hatano-Nelson Model in Cylinder Type}

\author{Yiming Zhao}
\affiliation{School of Science, Qingdao University of Technology, Shandong, China}
\author{Yazhuang Miao}
\affiliation{School of Science, Qingdao University of Technology, Shandong, China}
\author{Yihang Xing}
\affiliation{School of Science, Qingdao University of Technology, Shandong, China}
\author{Tianhui Qiu}
\affiliation{School of Science, Qingdao University of Technology, Shandong, China}
\author{Hongyang Ma$^\dagger$}
\affiliation{School of Science, Qingdao University of Technology, Shandong, China}
\email{hongyang_ma@aliyun.com}
\author{Xiaolong Zhao$^*$}
\affiliation{School of Science, Qingdao University of Technology, Shandong, China}
\email{zhaoxiaolong@qut.edu.cn}
\date{\today}

\begin{abstract}
Non-Abelian gauge offers a powerful route to engineer novel topological phenomena. Here, we systematically investigate a two-dimensional (2D) non-Hermitian Hatano-Nelson model incorporating SU(2) non-Abelian gauge, demonstrating the emergence of Hopf-link bulk braiding topology in the complex energy spectrum solely with $x$-direction nearest-neighbor couplings. Because of the limitations of exceptional point (EP) topology in fully capturing the rich non-Hermitian skin effect (NHSE) under non-Abelian influence, we introduce a novel polarization parameter derived from the generalized Brillouin zone (GBZ). This parameter quantitatively discerns left-, right-, and notably, bipolar skin modes, with its accuracy corroborated by directly encoding real-space eigenstate. Our findings reveal that non-Abelian gauge provides unprecedented influence over NHSE, compared with Abelian gauge and without gauge cases. Furthermore, we uncover unique characteristics of zero-imaginary-energy eigenstates at these topological boundaries, including pronounced degeneracy and bipolar localization which is unaffected by size effects, investigated via dynamical evolution and Inverse Participation Ratio (IPR). This work establishes a new paradigm for synthesizing and manipulating non-Hermitian topological phases driven by non-Abelian structures, opening avenues for topological engineering and holding promise for experimental realization in synthetic dimensional platforms.
\end{abstract}
\maketitle
\thispagestyle{fancy}
\lhead{}
\cfoot{}
\rfoot{}

\section{Introduction}

The exploration of non-Hermitian Hamiltonians has revolutionized our understanding of physical systems, unveiling a rich tapestry of phenomena that lack direct counterparts in conventional, closed Hermitian settings~\cite{AIP249435,makris2008beam,regensburger2012parity,feng2013experimental,doppler2016dynamically}. This burgeoning field has profoundly impacted diverse areas, from optics and photonics to condensed matter physics. Beyond the redefinition of conventional bulk topological invariants~\cite{AIP249435, wang2021topological,nature607271,yao2018edge,song2019non,prl076801,kunst2018biorthogonal,xiong2018does,nn1038, arcmp033133,prb195419}, non-Hermitian systems are distinguished by unique eigenvalue topologies in the complex energy plane~\cite{bergholtz2021exceptional,gong2018topological,kawabata2019symmetry,yokomizo2019non,zhang2020correspondence,okuma2020topological,yang2020non,wang2021topological,nature607271,hu2021knots,nn1038,borgnia2020non,prb195419}. Among these, the NHSE which is characterized by the anomalous accumulation of an extensive number of eigenstates at the system's boundaries, stands out as a hallmark phenomenon~\cite{lee2016anomalous,yao2018edge,xiong2018does,APX2109431,prl125186802,nc132496}. The profound implications of NHSE and other non-Hermitian topological features have spurred extensive theoretical research and have been experimentally validated across a multitude of physical platforms~\cite{weidemann2020topological,nn1038,wang2021generating,wang2021topological,nature607271,xiao2020non,zhang2021observation,zhang2021acoustic,wang2022non,ghatak2020observation,li2020critical,helbig2020generalized,hofmann2020reciprocal,liu2021non,zou2021observation}.

Synthetic gauge fields, encompassing both Abelian and more complex non-Abelian structures, have revolutionized the engineering of quantum matter, enabling the manipulation of topological properties. 
Seminal advancements have led to their realization in diverse physical platforms (e.g., ultracold atoms, electrical circuits and photonics), which have recently begun to be explored in non-Hermitian systems~\cite{RMP831523, NEWREFGOLDMAN, NEWREFYANGSCIENCE,lin2021steering,midya2018non,pla3831821,longhi2017non,longhi2015non,longhi2017nonadiabatic,wong2021topological,helbig2020generalized,prl070401}.
The non-Abelian gauge fields, into physical systems is known to induce profoundly new physics by introducing internal degrees of freedom and complex coupling dynamics~\cite{prl043804,apl9,nature63752}. 
Specifically, beyond the well-known requirement of long-range interactions for Hopf-link spectral topologies~\cite{nn1038,wang2021topological, hu2021knots,nature607271} non-Abelian gauge fields have been recently shown to bypass this constraint with only nearest-neighbor hopping\cite{prl043804}. 
Besides, recent studies have confirmed the existence of a specific localized behavior -- bipartite NHSE under long-range coupling~\cite{zhang2021acoustic,wang2021generating,prb045410}. So this raises a critical question: Can non-Abelian gauge fields induce bipartite NHSE even with only nearest-neighbor coupling too? Furthermore, the synergistic impact of conventional non-reciprocal transitions and non-Abelian gauge fields on NHSE remains to be explored.

In this work, we systematically investigate the impact of SU(2) non-Abelian gauge fields on the topological properties and NHSE in a 2D Hatano-Nelson model. 
We have demonstrated that, under specific conditions, the topological properties of the two-dimensional cylinder model can be characterized by the Hopf-link braiding degree. Furthermore, we systematically investigated the correspondence between different Hopf-link braiding degrees and the localized eigenstate behaviors.
Recognizing that EP analysis~\cite{jpmt444016}, while crucial for identifying phase transitions, often falls short in comprehensively characterizing the diverse phenomena of NHSE—particularly the emergence of complex localization patterns like bipolar skin modes—we turn to the GBZ framework for a more nuanced description. The GBZ not only accurately predicts the energy spectrum under open boundary conditions but also inherently encodes information about eigenstate localization. Indeed, it is well-established that the geometric features of the GBZ in the complex momentum plane, such as the extent to which its contours deviate from the unit circle and their relative positioning, directly correlate with the NHSE ~\cite{prl136802,yokomizo2019non}.

To specifically leverage these GBZ properties for quantitative analysis, we introduce a GBZ-based polarization parameter. This parameter is designed to quantitatively characterize the direction and extent of eigenstate localization, successfully distinguishing left-, right-, and notably, bipolar skin modes. 
The validity of this parameter is rigorously corroborated by direct real-space eigenstate distribution analysis. 
Finally, we delve into the behavior of zero-imaginary-energy eigenstates at topological phase boundaries, revealing their significant degeneracy and robust bipolar localization, which are further elucidated through dynamics evolution and IPR. Our findings not only unveil the rich physics arising from the synergy between non-Abelian gauge fields and non-Hermiticity but also provide a versatile framework for engineering complex topological phases and influencing boundary phenomena, with potential implications for topological waveguiding and robust signal manipulation in synthetic platforms.

This paper is organized as follows: Sec.\ref{Model} introduces the model. Sec.\ref{PhaseDiagram} details the phase diagram, braiding degree. Sec.\ref{sec:EERL} presents the GBZ-based polarization parameter for NHSE characterization and validated its validity by directly encoding the eigenstates. Sec.\ref{Energy} discusses eigenstate properties at phase boundary under the condition where neither the non-reciprocity of linear transition intensity nor the non-Abelian phase non-reciprocity is present. Sec.~\ref{CONC} provides conclusions. 

\begin{figure}[htbp]
    \centering
    \includegraphics[width=\linewidth]{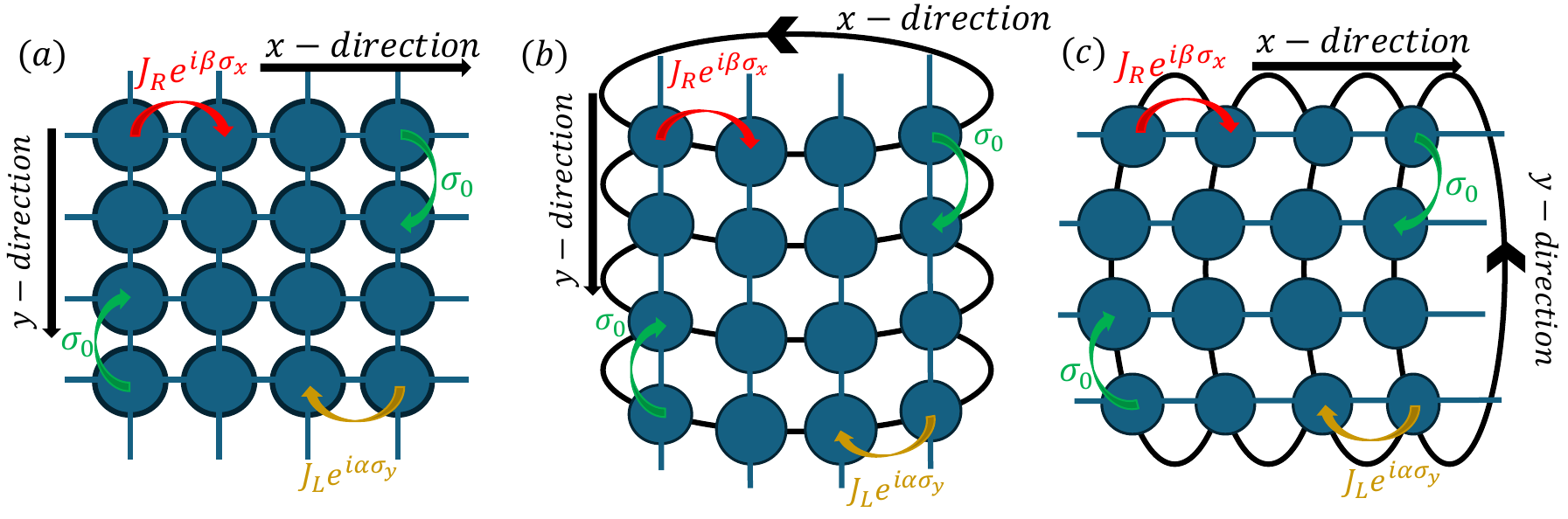}
    \caption{\textbf{Schematic drawing of the considered non-Abelian models.} 
    \textbf{(a)} 2D non-Abelian Hatano-Nelson model with non-reciprocality only in the $x$-direction under open boundary condition in both sides.
    \textbf{(b)-(c)} Model $(a)$ with one side under periodic boundary and showing cylindrical geometry.
    }
    \label{fig:model}
\end{figure}

\section{The model}
\label{Model}
The considered Hatano--Nelson model is a prototypical 2D system that demonstrates NHSE results from its nonreciprocal hoppings~\cite{hatano1996localization, arcmp033133,hatano1997}. 
In this work, we extend this model with SU(2) non-Abelian gauge, describing an excitation with spin-1/2 freedom on the lattice:
\begin{equation}
\label{eq:HN}
\begin{aligned}
    \mathbf{H} = \sum_{x,y}& \JL c^{\dagger}_{x,y}\mathrm{e}^{i A_a}c_{x+1,y} + \JR c^{\dagger}_{x+1,y}\mathrm{e}^{i A_b}c_{x,y}\\
    & + t(c^{\dagger}_{x,y+1}c_{x,y} + c^{\dagger}_{x,y}c_{x,y+1}),
\end{aligned}
\end{equation}
where $J_\mathrm{L(R)}$ is the real hopping amplitude leftward (rightward) in the $x$ directon, $t$ is the unit scale, namely, $t=1$. $c^{\dagger}_{x,y} (c_{x,y})$ is the creation (annihilation) operator at site $(x,y)$ with spin degree of freedom as $c^{\dagger}_{x,y} = \left(c^{\dagger}_{x,y\uparrow}, c^{\dagger}_{x,y\downarrow}\right)$, $c_{x,y} = \left(c_{x,y\uparrow},c_{x,y\downarrow}\right)^T$. 
$A_a = \alpha \sigma_a$, $A_b = \beta \sigma_b$ are the non-Abelian gauge components, where the Pauli matrices $\sigma_{a,b} (a,b \in (x,y,z))$ represent the introduction of spin degree of freedom and the gauge types $a,b$ determines whether the corresponding gauge is non-Abelian $(a \neq b)$ or not $(a = b)$. In this work, we mainly restrict ourselves to the non-Abelian case by considering the gauge types $a = y$ and $b = x$.
 $\alpha(\beta)$ is the corresponding nearest neighbour non-Abelian phases.
Notably, in Hamiltonian~\eqref{eq:HN}, both the hopping amplitudes $\JL$, $\JR$ and the non-Abelian gauge field components $A_a$, $A_b$ contribute to the non-Hermiticity. 
The Hamiltonian of corresponding models without gauge and under Abelian gauge are supplemented in Appendix A, which will be comparatively discussed in the following sections.

To investigate the topological phases in the cylinder models, which are acquired from the initial 2D model by Fourier transformation $\mathbf{c}_k^\dagger=\sum_\mathbf{i}e^{-i k\cdot\mathbf{i}}\mathbf{c_i^\dagger}$, and the corresponding Hamiltonian with periodic boundary condition (PBC) in $x$-direction and open boundary condition (OBC) in another dimension is:
\begin{equation} \mathbf{H}_\mathbf{k_x,y}=\begin{pmatrix}
\mathcal{H}_\mathbf{k_x,1}&\mathcal{H}_\mathbf{ k_x,2\rightarrow1}&~&0\\
\mathcal{H}_\mathbf{ k_x,1\rightarrow2}&\mathcal{H}_\mathbf{k_x,2}& \ddots&~\\
~&\ddots&\ddots&\mathcal{H}_\mathbf{ k_x,N\rightarrow N-1}\\
0&~&\mathcal{H}_\mathbf{k_x,N-1\rightarrow N}&\mathcal{H}_\mathbf{k_x,N}
\end{pmatrix},\end{equation}
which corresponds to Figure~\ref{fig:model} (b). Similarly, another cylinder model with PBC in $y$-direction and OBC in $x$-direction reads
\begin{equation}
\label{eq:realenergy}
 \mathbf{H}_\mathbf{x,k_y}=\begin{pmatrix}
\mathcal{H}_\mathbf{1,k_y}&\mathcal{H}_\mathbf{2\rightarrow 1,k_y}&~&0\\
\mathcal{H}_\mathbf{1\rightarrow 2,k_y}&\mathcal{H}_\mathbf{2,k_y}& \ddots&~\\
~&\ddots&\ddots&\mathcal{H}_\mathbf{N\rightarrow N-1,k_y}\\
0&~&\mathcal{H}_\mathbf{N-1\rightarrow N,k_y}&\mathcal{H}_\mathbf{N,k_y}
\end{pmatrix},\end{equation}
corresponding to Figure~\ref{fig:model} (c). Here $\mathcal{H}_\mathbf{k_x,y}=(J_L \cos{\alpha} e^{i k_x} + J_R \cos{\beta} e^{-i k_x}) \sigma_0 + i J_L \sin{\alpha} e^{i k_x}\sigma_y + i J_R \sin{\beta}  e^{-i k_x} \sigma_x$, $\mathcal{H}_\mathbf{k_x,y \rightarrow y+1}=\mathcal{H}_\mathbf{k_x,y \rightarrow y-1}=t \sigma_0$;
$\mathcal{H}_\mathbf{x,k_y}=2t \cos{k_y} \sigma_0$, $\mathcal{H}_\mathbf{x\rightarrow x-1,k_y }=J_L (\cos{\alpha} \sigma_0 + i \sin{\alpha} \sigma_y)$, $\mathcal{H}_\mathbf{x\rightarrow x+1,k_y }=J_R(\cos{\beta} \sigma_0 + i \sin{\beta} \sigma_x)$, and $\sigma_0$ is the identity matrix.

The Bloch Hamiltonian of Eq.~\eqref{eq:HN} with Fourier transformation in both $x$ and $y$-direction is
\begin{align}\label{eq:HN_nab_k}
     \mathbf{H}_{k_x,k_y} &= A(k_x,k_y)\sigma_0 + i \JR\sin{\beta}\mathrm{e}^{-i k_x}\sigma_x + i \JL\sin{\alpha}\mathrm{e}^{i k_x}\sigma_y,
\end{align}
where 
\begin{align}
    A(k_x,k_y) = \JL\cos{\alpha}\mathrm{e}^{i k_x} + \JR\cos{\beta}\mathrm{e}^{-i k_x} + 2t \cos{k_y},
\end{align}
and the eigen-energy of $  \mathcal{H}_{k_x,k_y}$ is given by
\begin{align} \label{eq:HN_nab_E}
    E_{\pm}(k_x,k_y) = A(k_x,k_y)\pm i\sqrt{\JR^2\sin^2{\beta}\mathrm{e}^{-i 2k_x} + \JL^2\sin^2{\alpha}\mathrm{e}^{i 2k_x}}.
\end{align}
The establishment of various Hamiltonian types and the calculation of eigen-energy have laid the foundation for the introduction of topological diagrams represented by the braiding degree in Sec.~\ref{PhaseDiagram} and behavior of corresponding eigenstates represented by a polarization parameter in Sec.~\ref{sec:EERL}.

\begin{figure}[htbp]
    \centering
    \includegraphics[width=\linewidth]{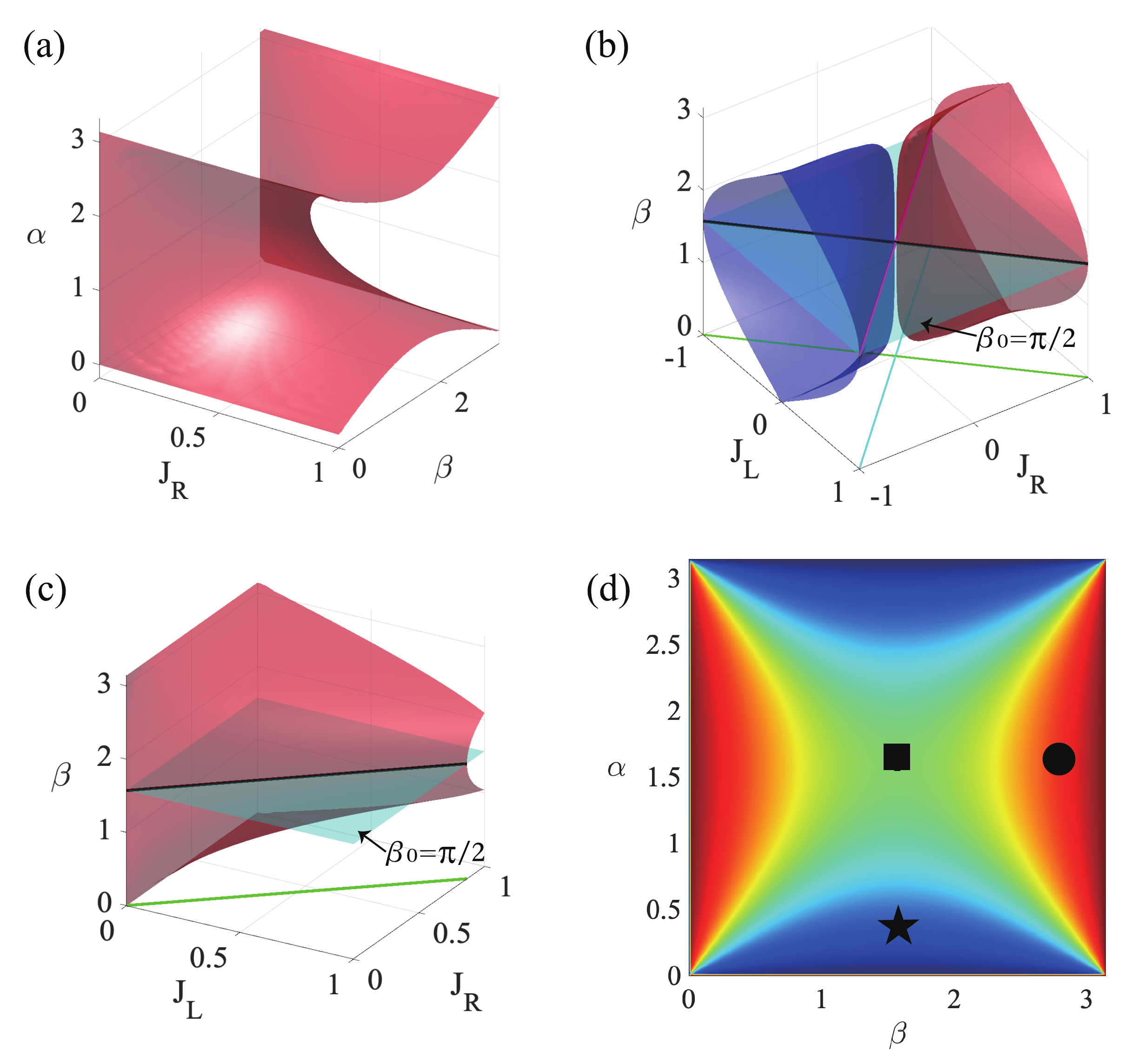}
    \caption{\textbf{Topological diagrams of the 2D Hatano-Nelson model with non-Abelian gauge in $x$-direction and ratio-quantitative index.} 
    \textbf{(a)} $\alpha-\beta-J_R$ phase diagram of the non-Abelian Hatano-Nelson model under $J_L$ = 0.5, describing two different phase regions split by the surface of EP.    %
    \textbf{(b)} The EP-surface in $\beta$-$J_L$-$J_R$ space with $\alpha = \pi/2$, $\beta \in (0,\pi)$, $J_L, J_R \in (-1,1)$ and the plane with $\beta = \pi/2$. 
    The surface intersects with the plane to form two intersection lines, which are projected onto the basal plane $\beta = 0$.  
    \textbf{(c)} The local magnification ($J_L, J_R \in (0,1)$) of subgraph (b) when $\alpha = \pi/3$ and the depiction of the intersection of plane $\beta = \pi/2$.
    \textbf{(d)} The graph described by a ratio-quantitative index~(\ref{eq:RQI}) with three sampling points marked by solid square, pentagram and solid circle.
    }
    \label{fig:EP_3D}
\end{figure}

\begin{figure}[htbp]
    \centering
    \includegraphics[width=\linewidth]{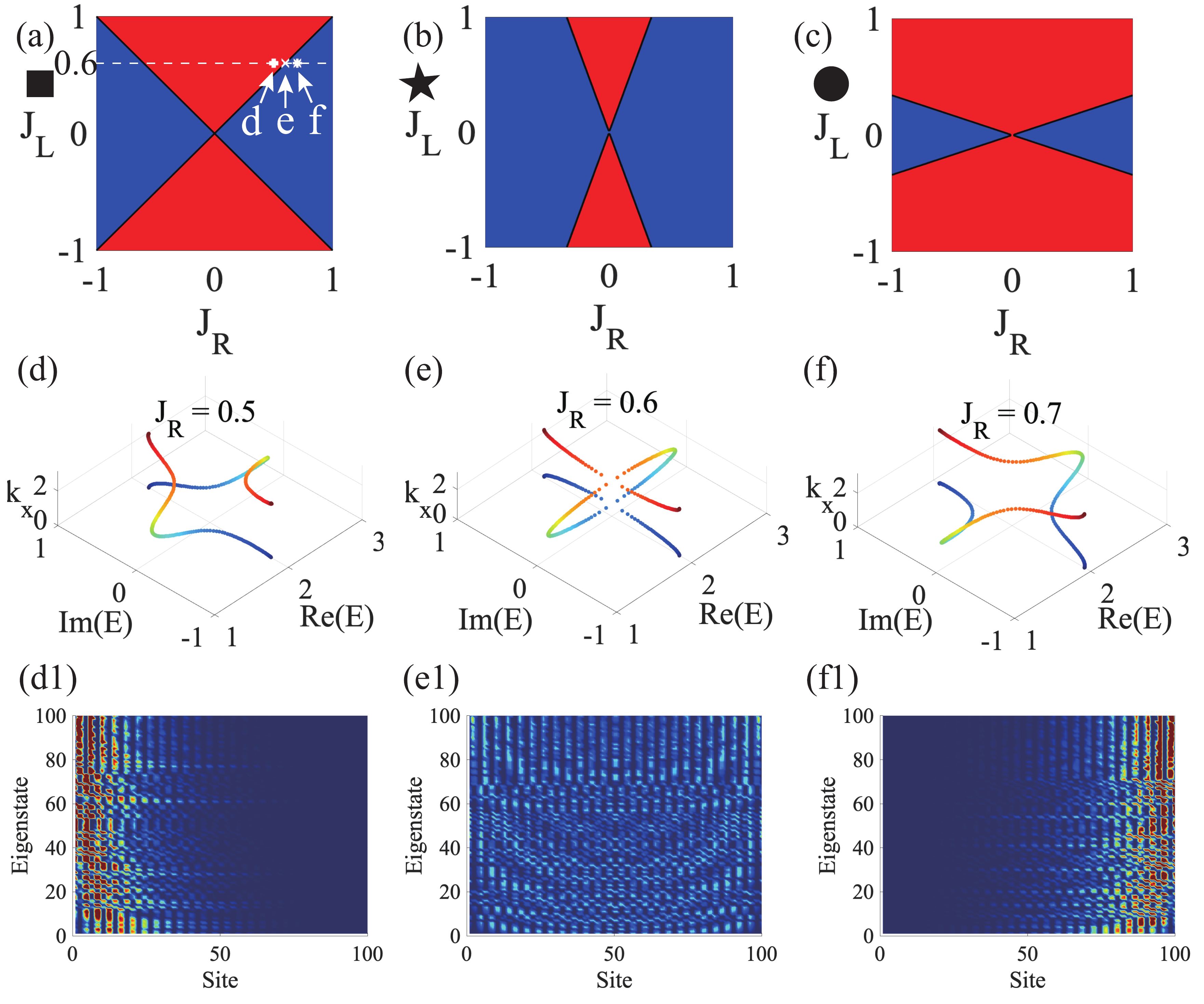}
    \caption{\textbf{Topological diagrams of the 2D Hatano-Nelson model with non-Abelian gauge in $x$-direction and the correspondence of eigenvalues and eigenstates in different phase regions.} 
    \textbf{(a)} $J_L$-$J_R$ phase diagram of the non-Abelian model in which two different phase regions split by the EP condition  Eq.~\eqref{eq:EP} corresponds to the subface projection in Figure~\ref{fig:EP_3D} (b). 
    Its corresponding phase parameters $\alpha$, $\beta$ refer to the solid square sampling point in Figure~\ref{fig:EP_3D} (d). 
    In this figure, the red area ($v$ = 2), phase boundary ($v$ = 0) and blue area ($v$ = -2) respectively have sampling points on the white line $J_L = 0.6$, which are indicated by plus sign $d$, crosse $e$ and asterisk $f$. %
    \textbf{(b)-(c)} $J_L$-$J_R$ phase diagrams with parameters $\alpha=\pi/9$, $\beta=\pi/2$ and $\alpha=\pi/2$, $\beta=8\pi/9$ correspond respectively to the pentagram and solid circle samples in Figure~\ref{fig:EP_3D} (d).
    \textbf{(d)-(f)} Hopf link corresponding to the three sampling points in (a).
    \textbf{(d1)-(f1)} Corresponding state density distribution of (d)-(f).
    }
    \label{fig:topozong1}
\end{figure}

\section{The Phase Diagram characterized by braiding degree and a ratio-quantitative index}\label{PhaseDiagram}
The dispersion in Eq.~\eqref{eq:HN_nab_E} forms Hopf link in ($\Re E$, $\Im E$, $k_x$) space and the phase transition of this energy braiding between two types of Hopf links is defined by a braiding degree~\cite{wang2021topological,nature607271,prl043804,nn1038} $\nu=\pm2$ (Figure~\ref{fig:topozong1} (a),(c)-(f)), where
\begin{align}
    \nu&\equiv\int_{0}^{2\pi} \frac{\diff k_x}{2\pi i}\frac{\diff}{\diff k_x}\mathrm{ln}\;\mathrm{det}\left(\mathcal{H}_{k_x,k_y} - \frac{1}{2}\mathrm{Tr}\mathcal{H}_{k_x,k_y}\right)\\
    &=\int_{0}^{2\pi} \frac{\diff k_x}{2\pi i}\frac{\diff}{\diff k_x}\mathrm{ln}\; (J^2_R \sin^2{\beta} e^{-i 2 k_x} + J^2_L \sin^2{\alpha} e^{i 2 k_x}),
\end{align} and the EP phase boundary is given by
\begin{equation}
{J^2_R} \sin^2{\beta}={J^2_L} \sin^2{\alpha}.
\label{eq:EP}
\end{equation}
In the three dimensional (3D) topological phase diagrams~\ref{fig:EP_3D} (a)-(c), different phase regions are separated by glossy dark red surface described by Eq.~\eqref{eq:EP}, and the surface satisfies $v=0$.
This EP phase boundary condition of this cylinder model is the same as the condition of the one-dimensional (1D) model with the Hamiltonian $\mathbf{H'} = \sum_{x} \JL c^{\dagger}_{x}\mathrm{e}^{i \alpha \sigma_y}c_{x+1} + \JR c^{\dagger}_{x+1}\mathrm{e}^{i \beta \sigma_x}c_{x}$ since $k_y$ is eliminated when calculating $\mathcal{H}_{k_x,k_y} - \frac{1}{2}\mathrm{Tr}\mathcal{H}_{k_x,k_y}$. 
In other words, this cylindrical model can be studied by analogy with a 1D model.

As Figure~\ref{fig:EP_3D} (b) shows, the EP phase boundary condition Eq.~\eqref{eq:EP} acts as two surfaces in $\beta-J_L-J_R$ space.
When $\beta$ takes different values $\beta_0$, that is, when the plane moves up and down, the angle between the two intersection lines will change. 
That means, the projected area of the region where the plane is inside the surface can be influenced by selection of $\alpha$ and $\beta$, and this region corresponds to one topological phase in $J_L$-$J_R$ diagrams as shown in Figure~\ref{fig:topozong1} (a).
In order to display the influence of different $\alpha$-$\beta$ selection on occupation of the phase areas in $J_L$-$J_R$ phase diagram, as Appendix B shows, we introduce an index $\mathcal{R}(\alpha,\beta)$ to quantify the ratio of $\nu = 2$ and $\nu = -2$ phase occupancy when $\alpha$ and $\beta$ vary.
As shown in Figure~\ref{fig:EP_3D} (d), by assigning different values to $\alpha$ and $\beta$ corresponding to the marks-square, pentagram and solid circle, the ratio of the area of the ‘red' phase ($\nu=2$) to the area of the ‘blue' phase ($\nu=-2$) when $J_L$ and $J_R$ take values within a certain range (Figure~\ref{fig:topozong1}(a)-(c)) can be characterized.
The redder the region is, the larger the region with topological number $\nu=2$ in the obtained topological phase diagram about $J_L$ and $J_R$ after taking this interval value as shown in Figure~\ref{fig:topozong1} (c). The same is true of the blue areas as Figure~\ref{fig:topozong1} (b) shows.
In Figure~\ref{fig:topozong1} (a)-(c), the phase transition criticality $v=0$ corresponds to the dividing lines of different color areas.

To explore the behavior of the eigenvalues and eigenstates in different topological regions, three parameter samples correspondence to three different phase numbers in Figure~\ref{fig:topozong1} (a), marked as d, e, f, respectively. 
The corresponding energy braiding is described in Figure~\ref{fig:topozong1} (d)-(f) and eigenstate density distribution are plotted in Figure~\ref{fig:topozong1} (d1)-(f1).
If the parameter selection ($J_L, J_R$) in Figure~\ref{fig:topozong1} (a) cross the phase variation boundary, the corresponding eigenenergy braiding will present different characteristics~\cite{video}.
In this situation,  
Figure~\ref{fig:topozong1} (d)-(f) confirm this transition, where different Hopf links corresponds to different sides of the EP phase critical line in Figure~\ref{fig:topozong1} (a)) ~cite{video}.
Besides,  Figure~\ref{fig:topozong1} (d1)-(f1) shows that the forms of the two energy braiding curves have a clear correspondence with the skin behavior of the eigenstate at the chain sites.
While non-Hermitian energy braiding exhibiting Hopf-link topology has been previously established, such demonstrations have generally necessitated longer-range hopping interactions~\cite{wang2021topological,hu2021knots}. 
Nevertheless, in this work the non-Abelian gauge enable the realization of the Hopf link using nearest-neighbor coupling only.

\section{Characterize NHSE with GBZ-based polarization parameter}\label{sec:EERL}
If non-Abelian gauge is introduced, the system will exhibit more intriguing skin effects including a bipolar skin effect even when $J_L = J_R$ and $\alpha = \beta$, i.e., when there is no non-reciprocity in both the hopping amplitude and non-Abelian phase, as shown in Appendix A. However, this kind of both-side localization cannot be fully described by the EP topological phase diagram.
Since the local radius of the GBZ is related to the direction of non-Hermitian skin~\cite{prl136802,yokomizo2019non}, so it is feasible to further explore the influence of non-Abelian gauge on the NHSE by designing a quantitative indicator based on GBZ. 
\subsection{GBZ for the non-Hermitian model}
To address the failure of conventional Bloch theory and bulk-boundary correspondence in non-Hermitian systems, GBZ is introduced to solve the problem~\cite{yao2018edge,yokomizo2019non}. In non-Hermitian systems, the conventional Bloch wave expansion with real quasi-momentum is inadequate so the complex momentum $\beta = e^{i k}$ is introduced. 
The eigenenergies $E$ are then obtained from the characteristic equation of the Hamiltonian \begin{equation}f(\beta,E)=\det[H(\beta)-E\mathbf{\sigma_0}]=0.\end{equation} 
By introducing the concept of auxiliary generalized Brillouin zone~\cite{yang2020non}, the following constraint equation can be obtained:
$\begin{cases}
f(\beta, E) = 0 \\
f(\beta e^{i\theta}, E) = 0,
\end{cases}$then a polynomial with respect to $\beta$ and $\theta$ can be obtained by canceling $E$ with resultant calculation:
$\mathcal{R}(E, \beta, \theta) = \text{Res}_E [f(\beta, E), f(\beta e^{i\theta}, E)] = 0$.
For a system whose characteristic equation is of order $2M$, denote the $2M$ solutions by $\beta_1,\beta_2,\ldots,\beta_{2M}$ which are ordered by their moduli: $|\beta_1|\leq|\beta_2|\leq\cdots\leq|\beta_{2M}|.$
The GBZ is then defined by the condition $|\beta_M|=|\beta_{M+1}|$, which selects the contour in the complex $\beta$-plane that balances the contributions of modes with decaying and growing amplitudes under open boundary conditions, and ensures that the modes included in the GBZ accurately capture the system's bulk properties.

\begin{figure}[htbp]
    \centering
    \includegraphics[width=\linewidth]{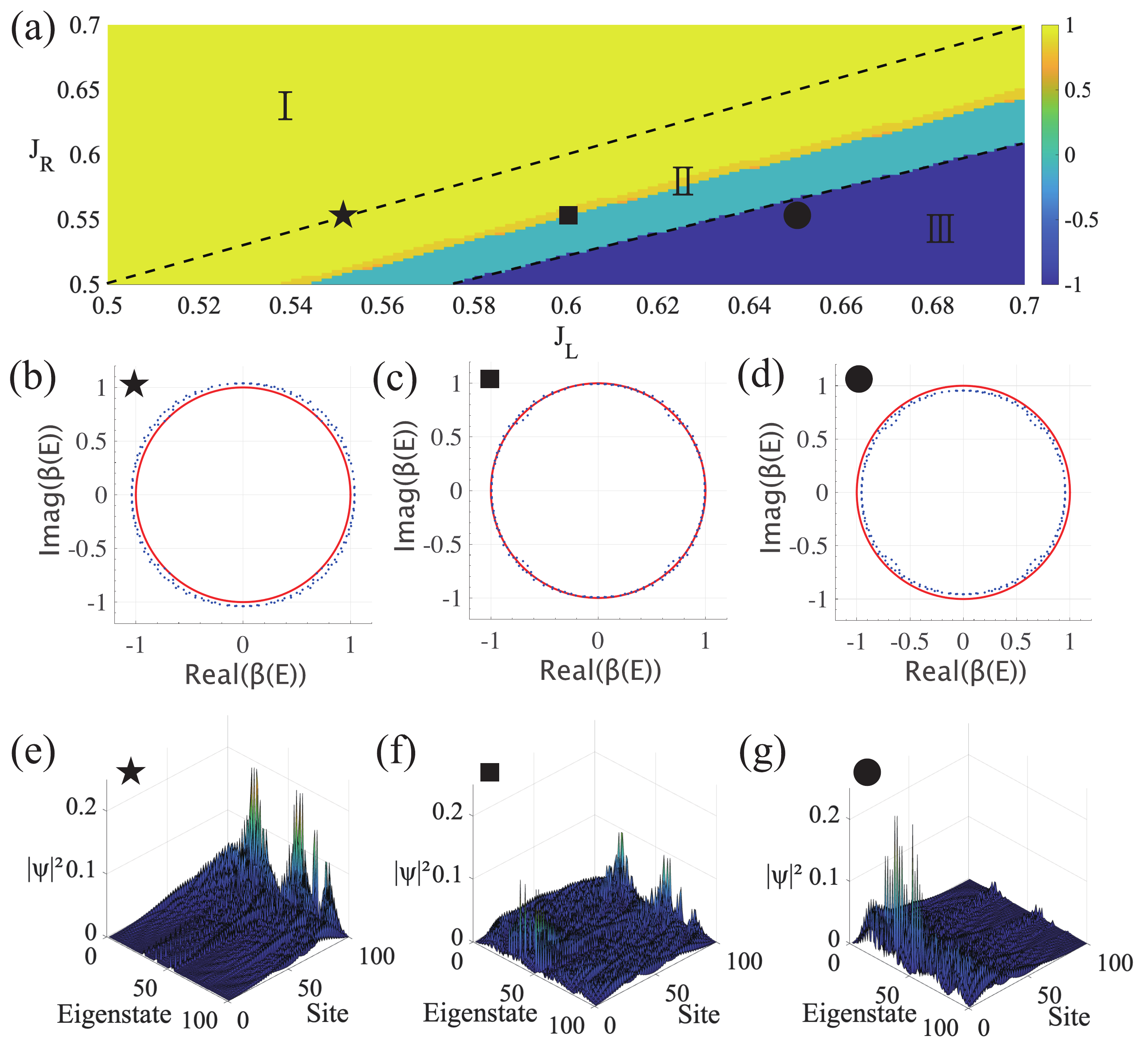}
    \caption{\textbf{Statistics of GBZ solutions and distribution of corresponding eigenstates of the 2D Hatano-Nelson model with designed polarization parameter in the $J_L$-$J_R$ plane.} 
    \textbf{(a)} Statistics of GBZ solutions of the non-Abelian model for fixed parameters $\alpha = \pi/3$, $\beta = \pi/2$ and $k_y = 1$, illustrating distinct phase regions characterized by the polarization parameter. The GBZ is calculated with respect to $k_x$.  %
    \textbf{(b)-(d)} GBZ corresponding to representative sampling points in (a), marked by pentagram, dot, and square, respectively.
    \textbf{(e)-(g)} Density of states distributions correspond to the same sampling points as in (b)-(d), consistent with the marks used in (a).
    }
    \label{fig:GBZjrjl}
\end{figure}

\subsection{GBZ-based polarization parameter}
We quantify the direction and degree of the localization by defining the polarization parameter $I_p$ based on the set of GBZ solutions $\{\beta_n\}_{n=1}^{2M}$:
\begin{equation}I_p=\frac{\mathcal{N}_>(R_d)-\mathcal{N}_<(R_d)}{\mathcal{N}_>(R_d)+\mathcal{N}_<(R_d)}\quad\in[-1,1],\label{IP}\end{equation}
where $R_d = 1$ is the radius of the Brillouin zone on the complex plane (Re($\beta$),Im($\beta$)).
$\mathcal{N}_>(R_d)=\sum_{n=1}^{2M}\Theta(|\beta_n|-R_d)$ signifies the cardinality of GBZ solutions whose modulus is bigger than $R_d$. $\Theta$ is Heaviside step function with $\Theta(\cdot) = \begin{cases} 
0, & \cdot < 0 \\
1, & \cdot \geq 0 
\end{cases}$. Similarly, $\mathcal{N}_<(R_d)=\sum_{n=1}^{2M}\Theta(R_d-|\beta_n|)$ represents the cardinality of the solutions whose modulus is smaller than $R_d$. 
When $|I_p|$ equals 1, it means that the outer boundary of the GBZ is all on the outside (inside) of the Brillouin zone, and the density distribution is completely skinned in one direction. In this situation, the state localization reaches the maximum of the parameter range. When $I_p \in (-1,1)$, it indicates that the skin effect on both sides is antagonistic, and the localization of the state gradually weakens as $I_p$ tends to 0.

For instance, we set $\alpha = \pi/3, \beta = \pi/2$ as $J_R$ and $J_L$ varies in [0.5,0.7] to explore the influence of a non-reciprocal non-Abelian gauge, as shown in Figure~\ref{fig:GBZjrjl}. If there is no non-reciprocal non-Abelian gauge, NHSE is completely affected by the imbalance between $J_L$ and $J_R$, and the phase region described by $I_p$ is divided diagonally, that is, there is a significant change in the skin direction near the diagonal $J_L = J_R$ as shown in Appendix A. However, because of the introduction of non-Abelian gauge with $\alpha \neq \beta$, the skin orientation within area II in Figure~\ref{fig:GBZjrjl} (a)(sandwiched between two dotted lines) has been changed, and the evaluation indicators $I_p$ clearly distinguish the areas with one-way skin (yellow area) and those with both-side skin (yellow-blue-middle area) in region II. 
Besides, the boundary between region II and region III coincides with the projection EP line in Figure~\ref{fig:EP_3D}(c).
The edges of different color regions in the Figure~\ref{fig:GBZjrjl} (a) exhibit distinct serrated patterns, and the central color band is divided into three distinct colors. This phenomenon fundamentally stems from insufficient computational precision. However, achieving this level of precision already requires a computational duration of 15 days.
The corresponding relationship between the evaluation indicator $I_p$ and the GBZ range is shown in the Figure~\ref{fig:GBZjrjl} (a)-(d), as the pentagram, square and circular sampling points are respectively located in the yellow, middle color and blue regions, which correspond to the lattice-end skin, the both-side skin and the lattice-head skin as Figure~\ref{fig:GBZjrjl} (e)-(g) shows. Furthermore, the influence of skin-orientation by the non-Abelian phases $\beta$ and $\alpha$ in the single hopping direction and the linear transition intensity $J_L$ on the skin effect were considered, as shown in Appendix C.

\begin{figure}[htbp]
    \centering
    \includegraphics[width=\linewidth]{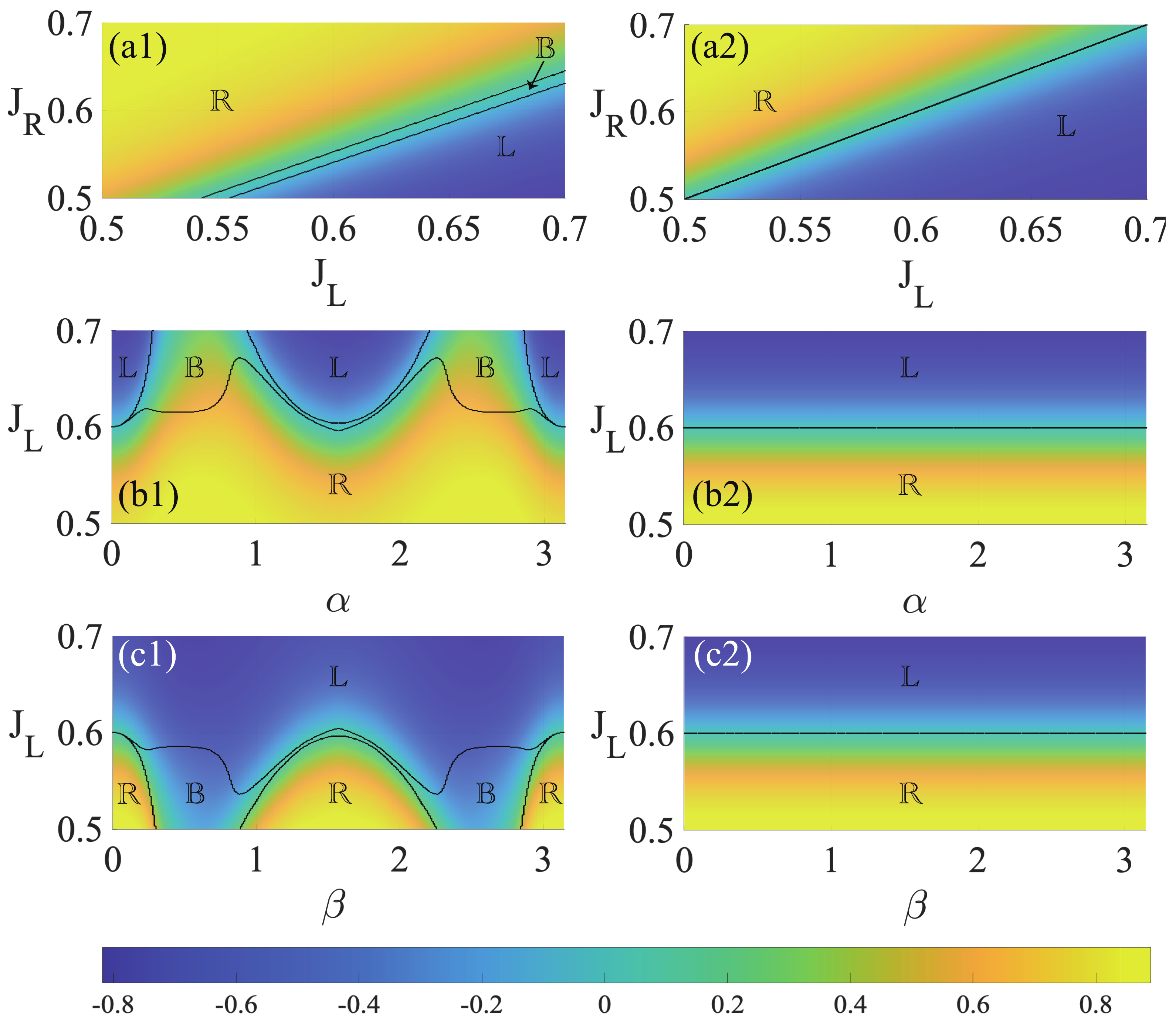}
    \caption{\textbf{The distribution of $\mathcal{C}(A,B)$ and $\mathcal{Q}$ acquired from the direct encoding of eigenvectors. The base color of these diagrams show the distribution of $\mathcal{C}(A,B)$ in Eq.~\eqref{cab}, and the regions divided by dashed lines are represented by ‘$\mathbb{B}$’, ‘$\mathbb{L}$’, ‘$\mathbb{R}$’.} 
    \textbf{(a1)} Phase diagram of the non-Abelian model for fixed parameters $\alpha = \pi/3$, $\beta = \pi/2$, illustrating distinct phase regions characterized by encoding eigenvectors.  %
    \textbf{(b1)} Eigenvector-based phase diagram with $J_R = 0.6$, $\beta = \pi/2$.
    \textbf{(c1)} Phase diagram based on encoded eigenvector with fixed parameters: $J_R = 0.6$, $\alpha = \pi/2$.
    \textbf{(a2)-(c2)} Comparison of (a1)-(c1) diagrams in the Abelian case, that is, $\sigma_a = \sigma_b = \sigma_y$ are set in Hamiltonian~\eqref{eq:HN}.
    $k_y = 1$ in all diagrams.
    }
    \label{fig:verification}
\end{figure}
\subsection{Verification by encoding eigenvectors}
The NHSE permits the accumulation of a multitude of eigenstates at the boundary, which stands in contrast to the Hermitian scenario where only a few boundary states are present. Therefore, it is feasible to investigate the localized behavior of NHSE by employing an averaging approach in the scale of eigenstate.
In order to further verify the rationality of the designed indicator $I_p$, and to directly probe the localization behavior from the real-space eigenstate distributions, we then compute a quantity $\mathcal{C}(A,B)$, which represents an expectation of these sites:
\begin{equation}\mathcal{C}(A,B)=\sum_{j=1}^{2N}\overline{P}(j)x_j,\label{cab}\end{equation}
where $(A,B)\in\{(J_R,J_L),(J_L,\alpha),(J_L,\beta)\}$ are selected variable combinations in this work. 
This calculation can be regarded as obtaining the total weight of $\overline{P}(j)$ on the $j$ scale.
$$\overline{P}(j)=\frac{1}{2N}\sum_{k=1}^{2N}|\psi_n(j)|^2, j=1,2,\ldots,2N,$$ 
is the average of probability density $|\psi_n(j)|^2$ for each eigenvector $\psi_n$, $n$ is the index of eigenvectors and $j$ indexes the eigenstates corresponding to the two spin-like degrees of freedom per site across the $N$ sites. 
In fact, this averaging operation can be regarded as a dimensionality reduction in the $k$ scale. 
$x_j=\frac{2(j-1)}{2N-1}-1,  j=1,2,\ldots,2N,$ is a linearly spaced array from -1 to 1 over $2N = 100$ points, which ensures $\mathcal{C}(A,B)$ can be uniformly encoded within the range of -1 to 1 according to different skin degrees. That is, when $\mathcal{C}(A,B) = -1$, it indicates that the skin degree towards the head point reaches the maximum, while when $\mathcal{C}(A,B) = 1$, it indicates that the skin degree towards the tail point reaches the maximum.

To provide a qualitative classification of the skin effect regimes, we further analyze the monotonicity of $\overline{P}(j)$. The system is first divided into two overall halves: the left half $\overline{P}_L=\{\overline{P}(j)|j=1,\ldots,N\}$ and the right half $\overline{P}_R=\{\overline{P}(j)|j=N+1,\ldots,2N\}$. Each of these halves is then further bisected to assess its internal trend. For a generic segment $S=\{s_1,s_2,\ldots,s_m\}$, we define its left sub-segment sum $S_{sub,L}$ and right sub-segment sum $S_{sub,R}$ as:
$S_{sub,L} = \sum_{i=1}^{m/2} s_i, \quad S_{sub,R} = \sum_{i=m/2 + 1}^{m} s_i,$
and the encoded monotonicity $M(S)$ of segment $S$ is then determined by comparing these sums:
\begin{equation}
M(S) = \begin{cases}
&1 , \text{if } S_{sub,R} > S_{sub,L} + \delta \\
&-1 , \text{if } S_{sub,R} < S_{sub,L} - \delta \\
&0 , \text{otherwise}
\end{cases},
\label{eq:M_S}
\end{equation}where $\delta$ is a small tolerance, and values 1, -1, 0 serve only to label distinct eigenstate localization patterns, with no implied magnitude meaning. Let $M_L=M(\overline{P}_L)$ and $M_R=M(\overline{P}_R)$ be the monotonicity values for the left and right halves of $\overline{P}(j)$, respectively. The overall skin effect regime is then secondarily encoded based on the pair $(M_L,M_R)$:
\begin{equation}
\mathcal{Q} = \begin{cases}
& 1, ~\text{if} (M_L, M_R) = (-1, 1) \text{ (Bipolar skin, ‘$\mathbb{B}$’)} \\
& 2, ~\text{if} (M_L, M_R) = (-1, -1) \text{ (Left-skin, ‘$\mathbb{L}$’)} \\
& 3, ~\text{if} (M_L, M_R) = (1, 1) \text{ (Right-skin, ‘$\mathbb{R}$’)} \\
& 4, ~\text{if} (M_L, M_R) = (0, 0) \text{ (Non-localized/flat, ‘$\mathbb{N}$’)} \\
& 5, ~\text{otherwise (Other combinations, ‘$\mathbb{O}$’)}.
\end{cases}
\label{eq:M_encode}
\end{equation}
The black lines overlaid on the surfaces in Figure~\ref{fig:verification} represent the boundaries between regions marked as ‘$\mathbb{B}$’, ‘$\mathbb{L}$’, ‘$\mathbb{R}$’, ‘$\mathbb{N}$’, $‘\mathbb{O}$’, associated with different $\mathcal{Q}$ values, thus delineating these qualitatively distinct skin effect regimes. Clearly, these phase diagrams in Figure~\ref{fig:verification} obtained through demonstrating the real-space eigenstate density directly (represented by the color map of $\mathcal{C}(A,B)$ and boundary lines of $\mathcal{Q}$ regions) is remarkably similar to the phase diagrams obtained using the designed polarization parameter $I_p$ (Figure~\ref{fig:GBZjrjl},~\ref{fig:GBZjlalpha},~\ref{fig:GBZjlbeta} (a)) which is rooted in the momentum-space GBZ formalism.

Furthermore, a comparative study between the non-Abelian and Abelian models reveals notable differences in their skin-effect: the non-Abelian model exhibits a shifted position of the non-unidirectional skin region compared to the Abelian counterpart by comparing Figure~\ref{fig:verification} (a1), (a2). 
This discrepancy arises from the fact that the skin effect in the model under Abelian gauge remains unaffected by nonreciprocal phase $\alpha \neq \beta$, whereas the model under non-Abelian gauge is sensitive to such phase combination as shown in Figure~\ref{fig:verification} (b1)-(c2)and Figure~\ref{fig:EP_33D} (b), (c).
Besides, all diagrams in Figure~\ref{fig:verification} exhibit no localized patterns ‘$\mathbb{N}$’ and ‘$\mathbb{O}$’ other than left-skin, right-skin and both-side skin.
While in the $J_L$-$\alpha$($\beta$) diagrams, namely, Figure~\ref{fig:verification} (b1)-(c2), the non-Abelian model demonstrates flexible adjustability of skin direction by monotonically tuning the phases $\alpha, \beta$ of the both hopping directions.
By directly utilizing $\mathcal{C}(A,B)$ and $\mathcal{Q}$ to characterize the NHSE of eigenstates in Figure~\ref{fig:verification}, one can provide validation that the designed polarization parameter $I_p$ holds significant value in quantifying the degree of eigenstate localization.

\section{The behavior of the eigenvalues and eigenstates for the critical boundary}\label{Energy}

\begin{figure}[htbp]
    \centering
    \includegraphics[width=\linewidth]{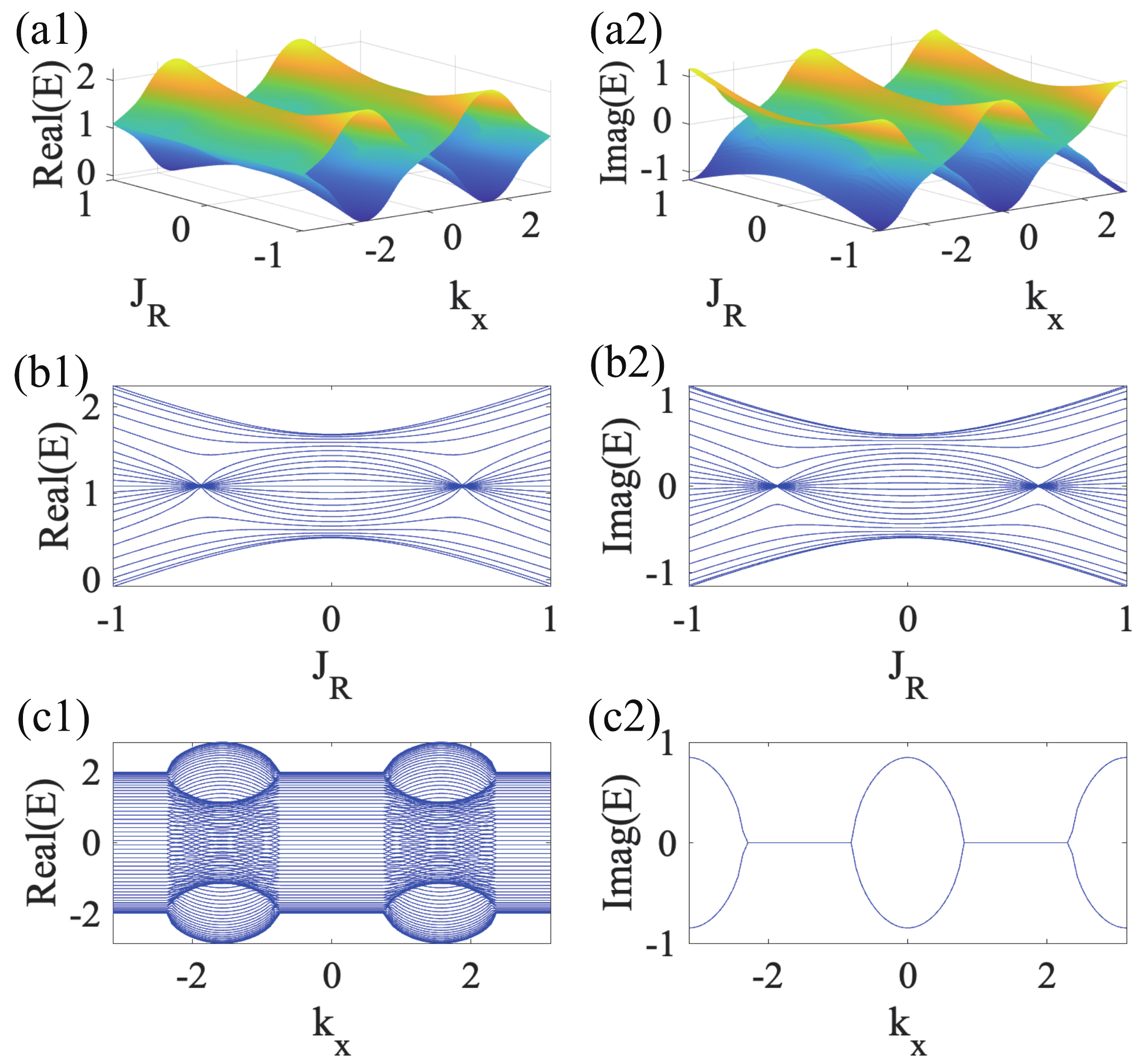}
    \caption{\textbf{The dispersion of the non-Abelian Hatano-Nelson model and the real space energy spectrum under PBC in both cylinder cases as shown in Figure~\ref{fig:model} (b), (c).} 
    \textbf{(a1)} The real part of the dispersion versus $J_R$ and $k_x$, with $k_y = 1$. %
    \textbf{(a2)} The imaginary part of the dispersion corresponding to (a1).
    \textbf{(b1)} The real part of the real space spectrum for model~(\ref{fig:model}) (c) versus $J_R$, and $k_y = 1$. The $x$-direction is under real PBC.
    \textbf{(b2)} The imaginary part of the real space spectrum corresponding to (b1).
    \textbf{(c1)} The real part of the real space spectrum for model~(\ref{fig:model}) (b) versus $k_x$, and $J_R = 0.6$. The $y$-direction is under real PBC.
    \textbf{(c2)} The imaginary part of the real space spectrum corresponding to (c1).
    All these subgraphs use the parameters: $J_L = 0.6$, $\alpha = \pi/2$, $\beta = \pi/2$.
    }
    \label{fig:energy}
\end{figure}

\begin{figure}[htbp]
    \centering
    \includegraphics[width=\linewidth]{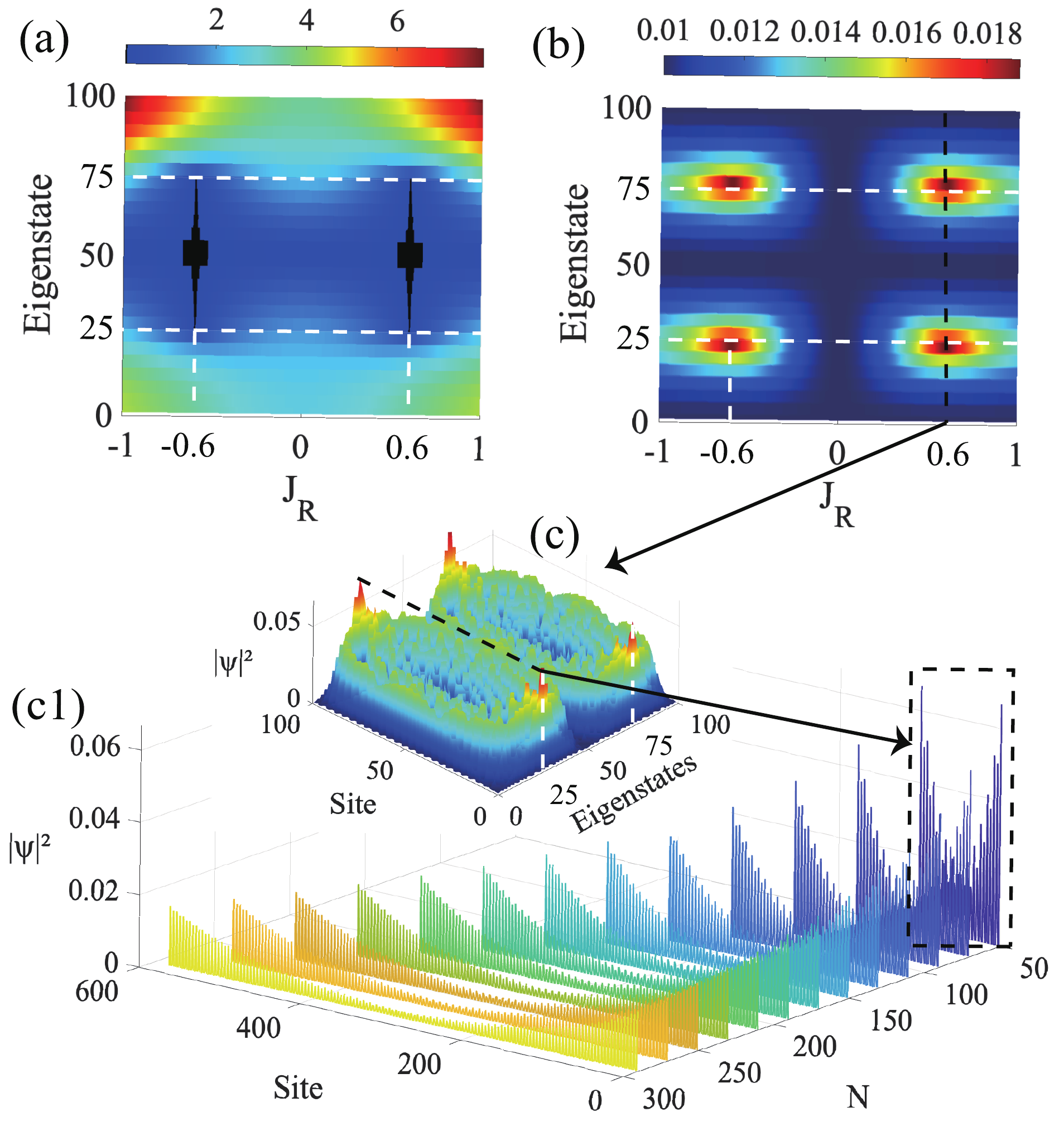}
    \caption{\textbf{The distribution of fidelity $\mathcal{D}_{n,J_R}$ from dynamics, IPR, and the DOS sampling at the critical boundary.
    } 
    \textbf{(a)} The $\mathcal{D}_{n,J_R}$ obtained from the dynamics of the $y$-direction cylinder model.
    \textbf{(b)} The distribution of IPR of the $y$-direction cylinder model with $t = 0$.
        \textbf{(c)} Probability density distribution $|\psi|^2$ of eigenstates for a system with unit number $N = 50$. Eigenstates are sorted by the imaginary part of the eigenvalues, and the parameters are the same to (a).
    \textbf{(c1)} Probability density distribution $|\psi|^2$ of the most edge-localized eigenstate for systems varying with size $N\in(50, 300)$. Fixed parameters are: $\alpha = \pi/2$, $\beta = \pi/2$, $J_R = 0.6$, $J_L = 0.6$, $k_y = 1$ in all subdiagrams.
    }
    \label{fig:trende0_1}
\end{figure}

The parameter regimes associated with distinct localization behaviors have been investigated as demonstrated in Sec.~\ref{sec:EERL} and supplemented in Appendix A, revealing that the introduction of the non-Abelian gauge gives rise to parameter regions exhibiting both-side skin effect near the critical boundary. Consequently, a deeper exploration of the eigenvalue and eigenstate characteristics at the critical boundary becomes particularly intriguing.


The sorted eigenvalues of the system under $y$-direction cylinder-types is shown in Figure~\ref{fig:energy} (b1), (b2).
When $|J_R| = J_L$ corresponding to the critical boundary, both the real and imaginary energy spectra of the cylindrical Hamiltonian exhibit significant aggregation, with the number of aggregated eigenstates accounting for half of the total. In contrast, for all other values of $J_R$, the eigen-energy levels accumulating at zero imaginary energy open up.
This can correspond to the side view of the momentum space energy band in the $E$-$J_R$ aspect in Figure~\ref{fig:energy} (a1), (a2). 
Additionally, in the real-energy dispersion relation of the $x$-direction cylinder model as Figure~\ref{fig:energy} (c1) shows, a periodic structural repetition emerges in the energy scale, which is directly attributed to the reciprocity of hopping without gauge along the $y$-direction. 
While the corresponding imaginary-energy dispersion relation Figure~\ref{fig:energy} (c2) shows extremely high degeneracy, where two distinct zero-energy regions are observed, corresponding to purely real eigenvalues.

\subsection{Dynamics reflects the topological phase transition of the system}\label{dynamics}

As seen from the preceding discussion, the topological phase transition of this model can manifest through intriguing characteristics of its complex eigenvalue spectrum in Figure~\ref{fig:trende0_1}. 
Since there exists a profound connection between the system's dynamics and the eigenvalue spectrum, we conjecture that the model's topology can be reflected through dynamics.
To further elucidate this reflection, we check the dynamics of the system with the initial states corresponding to the eigenvalues $E_{n,J_R} = E_{R,n,J_R} + i E_{I,n,J_R}$ shown in Figure~\ref{fig:energy} (b1), (b2). For each parameter set defined by $J_{R}$ (varied in the range $[-1, 1]$, with $J_{L} = 0.6$, $\alpha = \pi/2$, $\beta = \pi/2$, $t = 1$, $k_y = 1$, for $N = 50$ unit cells), the eigenstates $\psi_n(0)$ are first obtained and subsequently sorted according to the ascending value of their imaginary energy components.
Then the system evolves under the Schrödinger equation with the sorted eigenstates
\begin{equation}i\partial_t \psi_{n,J_R}(t) = H_{\text{PBC}} \psi_{n,J_R}(t),\label{eq:schr}\end{equation} up to a final time $T_{\text{final}}$, here $H_{\text{PBC}}$ is the Hamiltonian~(\ref{eq:realenergy}) under PBC in real space.
According to the Schrödinger equation~(\ref{eq:schr}), the real part $E_{R,n,J_R}$ acts as the change of phase, while the imaginary part $E_{I,n,J_R}$ relates to gain and loss. This means pure real energy spectrum does not leads to vanish or exponential scaling.

To quantify the cumulative change in the wavefunction's spatial amplitude profile due to non-zero imaginary energy components, a diagnostic measure $\mathcal{D}_{n,J_R}$ is defined as a function of eigenstate index $n$ and parameter $J_{R}$:
\begin{equation}
\mathcal{D}_{n,J_R} = \sum_j \left| \left|\psi_{n,J_R,j}(T_{\text{final}})\right| - \left|\psi_{n,J_R,j}(0)\right| \right|,
\label{eq:dynamical_diagnostic}
\end{equation}
where the sum is over all lattice sites, and $\left|\psi_{n,J_R,j}\right|$ means the absolute value of the amplitude of the wave function $\psi_{n,J_R,j}$'s component at lattice site $j$ with index $n$ of eigenstate and $J_R$.
This quantity $\mathcal{D}_{n,J_R}$ serves as not only the fidelity between $\psi_{n,J_R,j}(T_{\text{final}})$ and $\psi_{n,J_R,j}(0)$, but also a sensitive probe for the reality of the spectrum: if $E_{I,n,J_R} = 0$, then $\left|\psi_{n,J_R,j}(T_{\text{final}})\right| = \left|\psi_{n,J_R,j}(0)\right|$ for all sites, leading to $\mathcal{D}_{n,J_R} = 0$. While a non-vanish $E_{I,n,J_R}$ results in an exponential scaling of the wavefunction amplitudes.

As shown in Figure~\ref{fig:trende0_1}(a), regions characterized by $\mathcal{D}_{n,J_R} = 0$ directly correspond to eigenstates with purely real energies. 
Specifically, these dark regions appear when $|J_R| = J_L = 0.6$ where corresponds to the topological boundary in Figure~\ref{fig:topozong1} (a) as well as the degenerate points in Figure~\ref{fig:energy} (b2). Besides, the dark regions extend 50\% of the total eigenstates, consistent with the aggregation in the eigenvalue spectrum presented in Figure~\ref{fig:energy} (b2). 
To sum up, this method indicates the feasibility that the structure of the energy spectrum at the critical boundary can be presented through dynamics.

\subsection{IPR reflects the degeneracy of the energy bands for the system.}\label{IPR}

Complementary to the dynamics analysis in Subsec.~\ref{dynamics}, we investigate the spatial localization properties by employing the IPR for each normalized eigenstate $\psi_{n,J_R}$ obtained from the real-space Hamiltonian~(\ref{eq:realenergy}) under PBC:
\begin{equation}
\mathrm{IPR}_{n,J_R} = \sum_{j=1}^{L} |\psi_{n,J_R,j}|^4,
\label{eq:IPR_static}
\end{equation}
where $\psi_{n,J_R,j}$ is the amplitude of the $n$-th eigenstate on site $j$ with $J_R$. 
For each value of the coupling parameter $J_{R}$ (varied from -1 to 1 corresponding to the horizontal axis in Figure~\ref{fig:trende0_1}(b)), eigenstates are sorted according to the ascending imaginary part of eigenvalues $E_{I,n,J_R}$.
As Figure~\ref{fig:trende0_1}(b) presents, the IPR exhibits its highest degree of localization at the critical boundary.

When setting $J_R = J_L = 0.6$ and ordering the eigenstates by the imaginary part of their energy to plot the DOS, as shown in Figure~\ref{fig:trende0_1}(c), a bipolar skin effect emerges where index of eigenstate varying from 25 to 75, which signifies the boundary between pure-real and complex eigenenergies.
Notably, upon increasing the size of the system in Figure~\ref{fig:trende0_1}(c1), the most localized eigenstate consistently retains this bipolar skin effect, indicating that this localization is not size effect.
Together, IPR in Figure~\ref{fig:trende0_1} (b) and the distribution of DOS in Figure~\ref{fig:trende0_1} (c), (c1) provide a detailed picture of the eigenstate structure especially near the topological boundary.

The predicted localization behavior, particularly the pronounced bipolar characteristics at the critical boundary, holds significant potential for experimental observation in synthetic platforms, particularly in photonics \cite{weidemann2020topological}, quantum-engineered platforms \cite{xiao2020non}, and topolectrical circuits \cite{helbig2020generalized}, have enabled direct observation of phenomena like the pronounced boundary localization. Such experimental measurements would provide definitive evidence for the predicted bipolar skin effects and unique degeneracies stemming from non-Abelian gauge fields, as discussed herein.

\section{Conclusion}
\label{CONC}
This work has unveiled the profound impact of non-Abelian gauge on the topological characteristics and boundary phenomena within a non-Hermitian framework. By incorporating SU(2) gauge into a generalized 2D Hatano-Nelson model, we demonstrated the generation of Hopf-link bulk braiding topology in the complex energy spectrum and phase transitions.

Recognizing that EP analysis alone incompletely captures the rich NHSE phenomenology in the presence of non-Abelian fields, a GBZ-based polarization parameter $I_p$ was introduced. This metric quantitatively distinguishes the direction and extent of eigenstate localization, effectively discerning left-, right-, and significantly, both-side skin modes. 
The efficacy of $I_p$ was rigorously corroborated through direct analysis of real-space eigenstate distributions using an encoding scheme. 
The unprecedented impact of non-Abelian phases in our model towards the NHSE via the comparison with Abelian model has been discussed. Another key aspect of our investigation involves the distinct behavior of zero-imaginary-energy eigenstates at topological phase boundaries. These states were found to exhibit significant degeneracy and a robust bipolar localization, quantitatively assessed via IPR and their dynamics behavior.

Collectively, this work establishes non-Abelian gauge fields as a powerful tool for engineering complex non-Hermitian topological phases and diverse skin phenomena, significantly advancing the fundamental understanding of their intricate interplay. The demonstrated influence over topological structures and the NHSE holds considerable promise for experimental implementation in synthetic platforms like photonic lattices, opening avenues for advanced applications in topological waveguiding and quantum information science.


\section{Acknowledgements}
X. L. Zhao thanks National Natural Science Foundation of China (Grant No.12005110), Natural Science Foundation of Shandong Province (Grant No.ZR2020QA078, No.ZR2022QA110), Joint Fund of Natural Science Foundation of Shandong Province (Grant No.ZR2022LLZ012, No.ZR2021LLZ001), and Key R\&D Program of Shandong Province, China (Grant No.2023CXGC010901).

\section*{Appendix A: The state distribution under the common critical boundary of the three control groups: without gauge, under Abelian and non-Abelian gauge}
\label{sec:appendixA}
\begin{figure}[htbp]
    \centering
    \includegraphics[width=\linewidth]{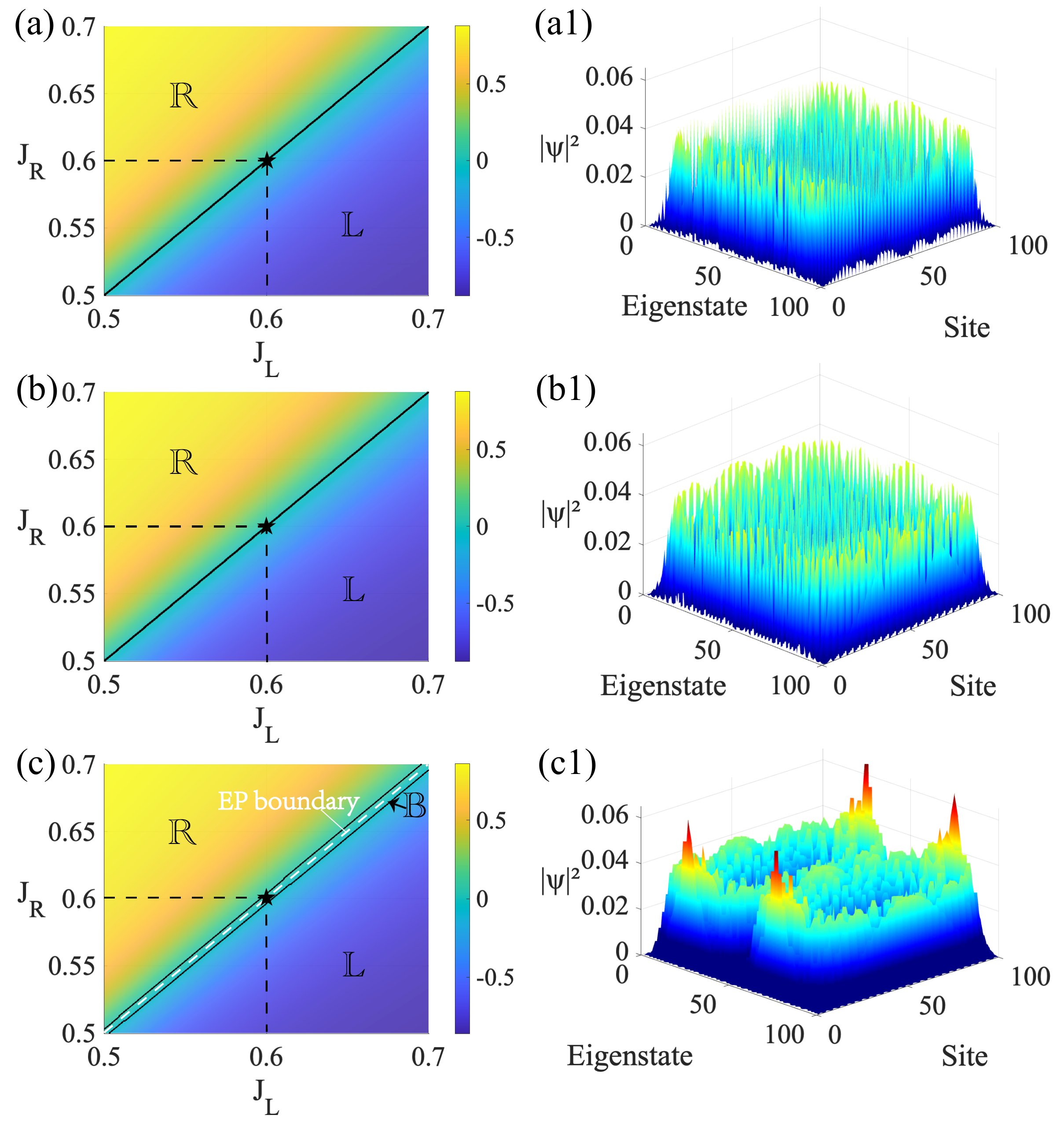}
    \caption{\textbf{The distribution of eigenstates for three 2D Hatano-Nelson models ~(\ref{eq:withoutNAHN}),~(\ref{eq:AHN}) and~(\ref{eq:HN}) in $y$-direction cylinder type with the same parameters, and their corresponding in the $\mathcal{C}(A,B)$ diagrams.} 
    \textbf{(a)} $\mathcal{C}(A,B)$ diagram for the model~(\ref{eq:withoutNAHN}) without gauge. %
    \textbf{(a1)}Eigenstate distributions of the model without gauge with $J_L = J_R =0.6$ corresponding to the pentagram sample in (a).
    \textbf{(b)} $\mathcal{C}(A,B)$ diagram for the Abelian model~(\ref{eq:AHN}) under $\alpha$ = $\pi$/2, $\beta$ = $\pi$/2.  
    \textbf{(b1)} Eigenstate distributions of the model under Abelian gauge with $J_L = J_R =0.6$ corresponding to the pentagram sample in (b).
    \textbf{(c)} $\mathcal{C}(A,B)$ diagram for the  non-Abelian model~(\ref{eq:HN}) under $\alpha$ = $\pi$/2, $\beta$ = $\pi$/2. %
    \textbf{(c1)} Eigenstate distributions of the model under non-Abelian gauge with $J_L = J_R =0.6$ corresponding to the pentagram sample in (c).
    }
    \label{fig:EP_33D}
\end{figure}

In this task, two models are set as control groups of investigated non-Abelian model, the one without gauge is:
\begin{equation}
\label{eq:withoutNAHN}
\begin{aligned}
    \mathbf{H} = \sum_{x,y}& \JL c^{\dagger}_{x,y} c_{x+1,y} + \JR c^{\dagger}_{x+1,y} c_{x,y} + t(c^{\dagger}_{x,y+1}c_{x,y} + c^{\dagger}_{x,y}c_{x,y+1})
\end{aligned}
\end{equation} with $y$-direction cylinder Hamiltonian form is the same as Hamiltonian~(\ref{eq:realenergy}), where $\mathcal{H}_\mathbf{x,k_y}=2t \cos{k_y} \sigma_0$, $\mathcal{H}_\mathbf{x\rightarrow x-1,k_y }=J_L \sigma_0$, $\mathcal{H}_\mathbf{x\rightarrow x+1,k_y }=J_R \sigma_0$.
Besides, another control model with Abelian gauge is:
\begin{equation}
\label{eq:AHN}
\begin{aligned}
    \mathbf{H} = \sum_{x,y}& \JL c^{\dagger}_{x,y}\mathrm{e}^{i \alpha \sigma_y}c_{x+1,y} + \JR c^{\dagger}_{x+1,y}\mathrm{e}^{i \beta \sigma_y}c_{x,y}\\
    & + t(c^{\dagger}_{x,y+1}c_{x,y} + c^{\dagger}_{x,y}c_{x,y+1})
\end{aligned}
\end{equation} with $y$-direction cylinder Hamiltonian form is the same as Hamiltonian~(\ref{eq:realenergy}), where $\mathcal{H}_\mathbf{x,k_y}=2t \cos{k_y} \sigma_0$, $\mathcal{H}_\mathbf{x\rightarrow x-1,k_y}=J_L (\cos{\alpha} \sigma_0 + i \sin{\alpha} \sigma_y)$, $\mathcal{H}_\mathbf{x\rightarrow x+1,k_y }=J_R(\cos{\beta} \sigma_0 + i \sin{\beta} \sigma_y)$. The two models are treated as a control group of model~(\ref{eq:HN}) with non-Abelian gauge.

A comparative analysis of the eigenstate distributions at the common parameter point across the three models reveals a distinct bipolar skin effect emerging specifically in the non-Abelian model, which fundamentally originates from the selected non-Abelian SU(2) generators $\sigma_y$, $\sigma_x$.
Moreover, as shown in Figure~\ref{fig:EP_33D} (c), since the transition of the skin effect in the system—from one side to the other—always involves a gradual change near the EP phase boundary, the bipolar localization effect observed in Figure~\ref{fig:EP_33D} (c1) on the EP boundary (red dotted line) persists within a certain parameter range in the region $‘\mathbb{B}’$.

\section*{Appendix B: Ratio-quantitative index}
\label{sec:RQI}
The ratio-quantitative index is defined as:
\begin{equation}\begin{aligned}
\mathcal{R}(\alpha,\beta)& =
\begin{cases}
1, & \mathrm{if~}\sin\beta=0 ~and ~\sin\alpha>0, \\
0, & \mathrm{if~}\sin\alpha=0 ~or~ \sin\beta=\sin\alpha=0, \\
1-\frac{\sin\alpha}{2\sin\beta}, & \mathrm{if~}0<\sin\alpha\leq\sin\beta, \\
\frac{\sin\beta}{2 \sin\alpha}, & \mathrm{if~}\sin\alpha>\sin\beta>0
\end{cases}
\end{aligned}\label{eq:RQI}\end{equation} 
corresponding to Figure~\ref{fig:EP_3D} (d).
The calculation is based on the EP phase boundary Eq.~\eqref{eq:EP}, which quantifies the proportion of $\nu = 2$ phase in $J_L-J_R$ diagrams~\ref{fig:EP_3D} (a)-(c) under different selections of non-Abelian phases $\alpha$, $\beta$.

\section*{Appendix C: Detailed Exploration of NHSE Tunability via two Non-Abelian Phases and Hopping Amplitude $J_L$}
\label{sec:appendixB}

This part provides a more detailed exploration of how the NHSE can be tuned by varying specific non-Abelian phase parameters ($\alpha$, $\beta$) in conjunction with the hopping amplitude $J_L$. The phase diagrams presented here are constructed using the polarization parameter $I_p$ defined in Sec.~\ref{sec:EERL}, further illustrating its utility in capturing the nuanced localization behavior of eigenstates. For all subplots in this appendix, the transverse momentum is fixed at $k_y=1$, and the system size remains $N=50$.

\subsection{Interplay between Hopping Amplitude $J_L$ and Non-Abelian Phase $\alpha$}

\begin{figure}[htbp]
    \centering
    \includegraphics[width=\linewidth]{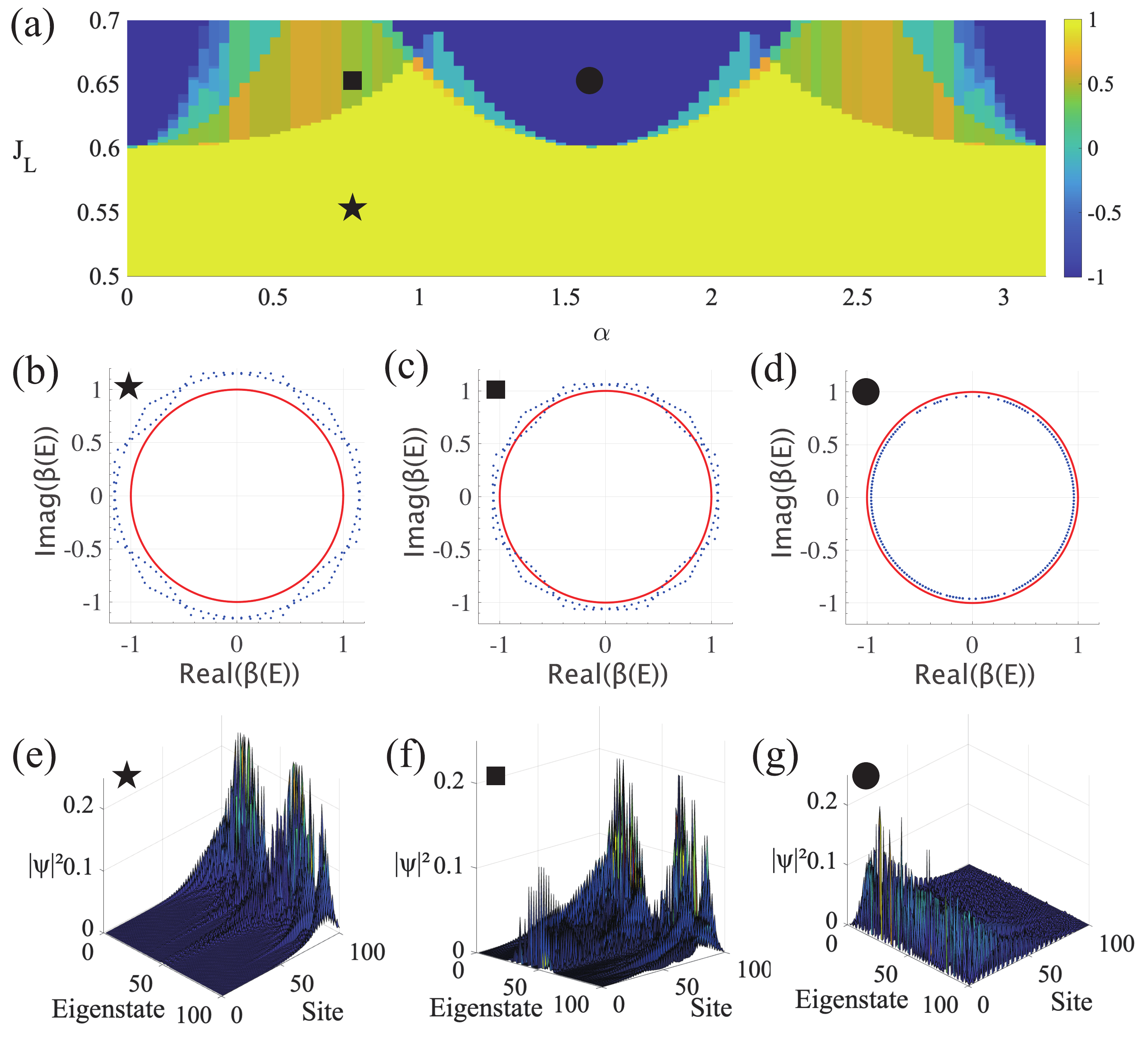}
    \caption{\textbf{Statistics of GBZ solutions and distribution of corresponding eigenstates of the 2D Hatano-Nelson model with designed polarization parameter~(\ref{IP}) reflecting the relationship between $J_L$ and $\alpha$.} 
    \textbf{(a)} Statistics of GBZ solutions of the non-Abelian model under $J_R = 0.6$, $\beta = \pi/2$ and $k_y = 1$, describing different phase spaces represented by polarization parameter $I_p$. The GBZ is calculated with respect to $k_x$. %
    \textbf{(b)-(d)} GBZ corresponding to different sampling points in (a), and the samples are distinguished by pentagrams, dots and squares.
    \textbf{(e)-(g)} The density of states distribution corresponding to different sampling points in (a) are the same as those in (b)-(d).
    }
    \label{fig:GBZjlalpha}
\end{figure}

\begin{figure}[htbp]
    \centering
    \includegraphics[width=\linewidth]{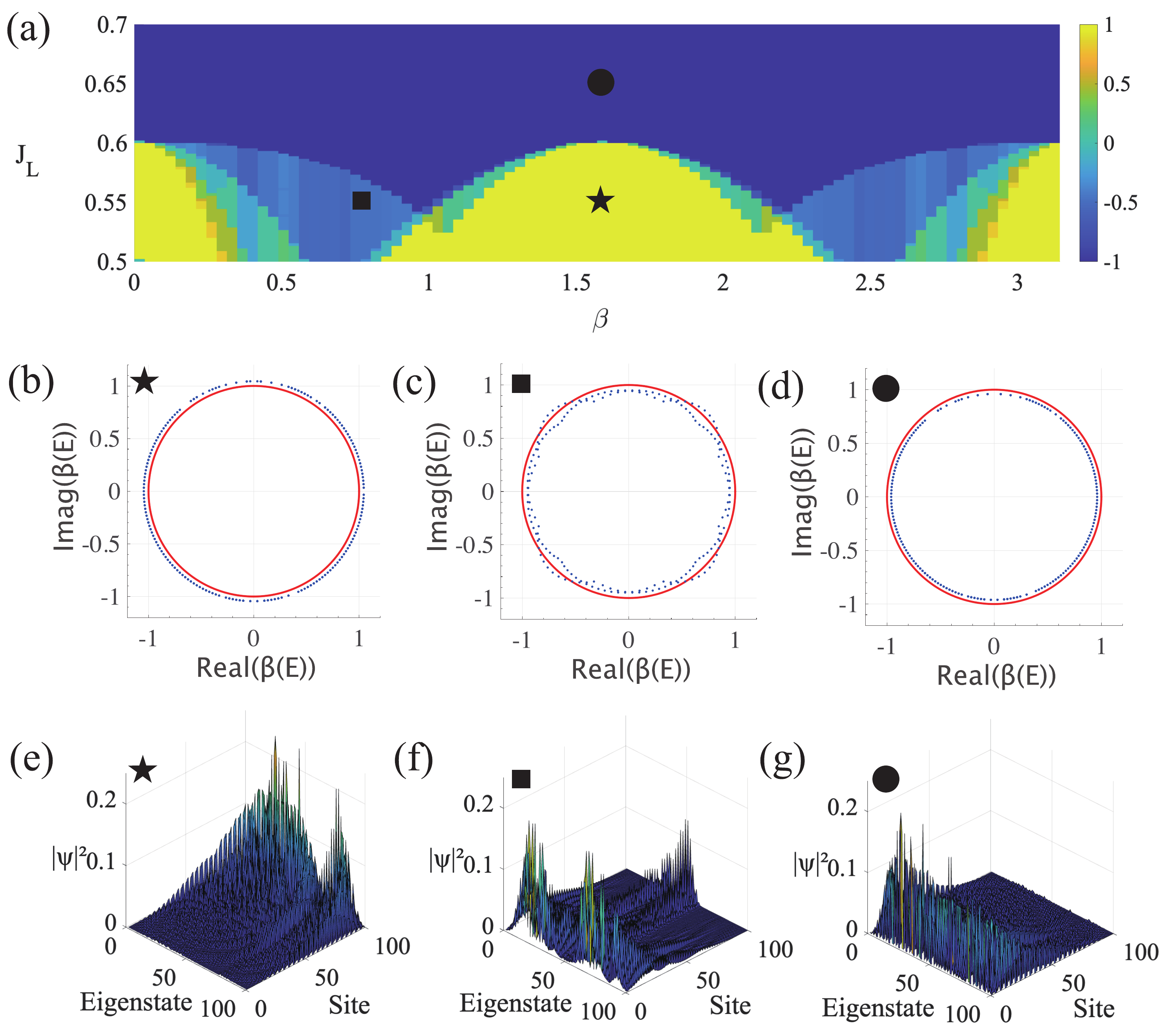}
    \caption{\textbf{Statistics of GBZ solutions and distribution of corresponding eigenstates of 2D Hatano-Nelson model with designed polarization parameter~(\ref{IP}) reflecting the relationship between $J_L$ and $\beta$} 
    \textbf{(a)} Statistics of GBZ solutions of the non-Abelian model under $J_R = 0.6$, $\alpha = \pi/2$ and $k_y = 1$, depicting distinct phase regions characterized by the polarization parameter $I_p$. The GBZ is calculated with respect to $k_x$.%
    \textbf{(b)-(d)} GBZ corresponding to representative sampling points in (a), denoted by pentagrams, dots, and squares.
    \textbf{(e)-(g)} Density of states distributions associated with the same sampling points in (b)-(d) are marked in the same way as those in (a).
    }
    \label{fig:GBZjlbeta}
\end{figure}

As shown in Figure~\ref{fig:GBZjlalpha} (a) which is mapped by the polarization parameter $I_p$, the NHSE characteristics as a function of the leftward hopping amplitude $J_L$ and the non-Abelian phase $\alpha$ associated with this hopping, while keeping the rightward hopping parameters fixed ($J_R = 0.6$, $\beta = \pi/2$).
It highlights a competitive interplay: if $J_L > J_R$, monotonous change of $\alpha$ can neatly alter the existing location mode, leading to rotation of left- ($I_p \approx  0$), right- ($I_p \approx  0$), or both-side ($-1 < I_p < 1$) skin effects.

The representative GBZs in Figure~\ref{fig:GBZjlalpha} (b)-(d) for selected parameter points (pentagram, square, circle in Figure~\ref{fig:GBZjlalpha} (a)) visually correspond to $I_p$. The shape and extent of the GBZ directly correlate with the skin direction: a GBZ extending predominantly outside (inside) the unit circle in the complex $\beta$-plane corresponds to rightward (leftward) localization. The corresponding eigenstate density distributions in Figure~\ref{fig:GBZjlalpha} (e)-(g) confirm these localization correspondences. Notably, the perfect match between Figure~\ref{fig:GBZjlalpha} (a) and Figure~\ref{fig:verification} (b1) described by $C(J_L,\alpha)$ underscores the feasibility of the polarization parameter $I_p$ in capturing the skin effect driven by the interplay of hopping amplitude and its associated non-Abelian phase in the same hopping direction.

\subsection{Interplay between Hopping Amplitude $J_L$ and Non-Abelian Phase $\beta$}

Figure~\ref{fig:GBZjlbeta} explores a different case of non-Abelian influence, examining the NHSE as a function of $J_L$ and the non-Abelian phase $\beta$ associated with the opposite hopping, with $J_R = 0.6$ and $\alpha = \pi/2$ (associated with $J_L$) held constant.
The polarization parameter map Figure~\ref{fig:GBZjlbeta} (a) demonstrates that the non-Abelian phase $\beta$ of the opposing hopping term also exerts significant influence on the skin localization when $J_L < J_R$. This also showcases a ‘competitive' or ‘cooperative' effect, where monotonous change of $\beta$ can either enhance or counteract the skin tendency induced by the $J_L/J_R$ imbalance and another phase $\alpha$. 
This behavior is rooted in how the non-Abelian phases modify the effective non-reciprocity of the system, since the effective non-reciprocity is jointly determined by the non-reciprocity of the linear hopping amplitudes and the non-reciprocity of the non-Abelian phases after introducing the non-Abelian gauge. Besides, the GBZs in Figure~\ref{fig:GBZjlbeta} (b)-(d) corresponds to the sampling points in Figure~\ref{fig:GBZjlbeta} (a). It is worth noting that, as Figure~\ref{fig:GBZjlbeta} (c) shows, the GBZ boundary corresponding to the middle color area in Figure~\ref{fig:GBZjlbeta} (a) is no longer a regular circle. The fundamental cause for such intricate GBZ shapes, leading to the yellow-blue-middle color areas indicating both-side and transitional skin effects, is the way non-Abelian gauge can cause the GBZ boundary to stretch, distort. The corresponding DOS in Figure~\ref{fig:GBZjlbeta} (e)-(g) once again confirm the value of $I_p$ in reflecting NHSE.

Collectively, the polarization parameter $I_p$ serves as an effective quantitative tool for mapping and understanding these complex skin modes, which arise from the intricate interplay between hopping amplitudes and their co-directional and counter-directional non-Abelian phases.

\end{document}